\documentclass[aps,prc,twocolumn,superscriptaddress]{revtex4-1}
\usepackage{graphicx}
\usepackage{amsmath}
\usepackage{subfigure}
\usepackage{subfigure,dcolumn}
\usepackage{appendix}
\usepackage{xcolor}
\usepackage{ulem}
\usepackage[colorlinks,
linkcolor=blue,
anchorcolor=blue,
urlcolor=red,
citecolor=blue]{hyperref}
\begin{document}
	
	
\title{Effects of in-medium $NN$ inelastic cross sections and the high-momentum tail of nucleon momentum distributions on pion production in heavy-ion collisions}
	

\author{Pengcheng Li}
\affiliation{School of Science, Huzhou Normal University, 313000 Huzhou, China}
\affiliation{Bogoliubov Laboratory of Theoretical Physics, Joint Institute for Nuclear Research, 141980 Dubna, Russia}
\author{A. B. Larionov}
\affiliation{Bogoliubov Laboratory of Theoretical Physics, Joint Institute for Nuclear Research, 141980 Dubna, Russia}
\author{Yongjia Wang}
\affiliation{School of Science, Huzhou Normal University, 313000 Huzhou, China}
\author{Gaochan Yong}
\affiliation{Institute of Modern Physics, Chinese Academy of Sciences, 730000 Lanzhou, China}
\author{Qingfeng Li}
\email[Corresponding author, ]{liqf@huznu.edu.cn}
\affiliation{School of Science, Huzhou Normal University, 313000 Huzhou, China}

\date{\today}

\begin{abstract}		

Pion production in intermediate-energy heavy-ion collisions (HICs) provides a sensitive probe of the nuclear equation of state and of the isospin dependence of reaction dynamics.
In particular, pion production near threshold is strongly affected by the nucleon-nucleon ($NN$) inelastic cross sections and by the high-momentum components of the nucleon momentum distribution.
To explore the influence of these two ingredients on pion production and charged-pion ratios, the in-medium $NN$ inelastic cross sections calculated within the relativistic Boltzmann-Uehling-Uhlenbeck transport theory and the short-range correlation induced high-momentum tail (HMT) are introduced into the Ultra-relativistic Quantum Molecular Dynamics model.
By simulating Au+Au collisions at intermediate energies, we find that the in-medium modification of the $NN$ inelastic cross sections suppresses the pion multiplicity by reducing the probability of $N\Delta$ production in dense matter.
The HMT, on the other hand, enhances the high-momentum components of nucleons and modifies the available energy in individual $NN$ collisions, thereby affecting $NN\rightarrow N\Delta$ reactions and the subsequent pion production. 
With the simultaneous inclusion of these two effects, the pion yields and the $\pi^-/\pi^+$ ratio data can be better reproduced. 
These results highlight the importance of both the in-medium reaction cross sections and short-range correlation induced high-momentum components in pion production in intermediate energy HICs.

\end{abstract}

\pacs{}

\maketitle
	
\section{Introduction}\label{section1}

The properties of nuclear matter under extreme conditions are among the central topics in nuclear physics and nuclear astrophysics~\cite{Huth:2021bsp,Tsang:2023vhh}.
Intermediate-energy heavy-ion collisions (HICs) provide a unique opportunity to create compressed baryonic matter in the laboratory and to investigate the nuclear equation of state (EoS), especially its density dependence at suprasaturation densities~\cite{Danielewicz:2002pu,Sorensen:2023zkk}.
Such information is closely related to the structure of nuclei, the dynamics of core-collapse supernovae, and the properties of neutron stars~\cite{Niksic:2011sg,Dietrich:2020efo}.
Among various experimental observables, pion production has long been regarded as a sensitive probe of the high-density stage of intermediate-energy HICs, because pions are mainly produced through inelastic baryon-baryon reactions such as $NN\rightarrow N\Delta$, followed by $\Delta\rightarrow N\pi$ decay, and are therefore closely connected with the compression dynamics and the isospin-dependent reaction mechanism in dense nuclear matter~\cite{Harris:1984up,Li:2005gfa}.

In neutron-rich systems, the charged-pion ratio $\pi^-/\pi^+$ has attracted particular attention as a probe of the nuclear symmetry energy at high densities~\cite{Li:2008gp}.
Since $\pi^-$ and $\pi^+$ are dominantly related to neutron-neutron and proton-proton collision channels, respectively, the $\pi^-/\pi^+$ ratio is expected to carry information on the neutron-to-proton composition of the high-density region.
Experimental measurements of pion multiplicities, spectra, and charged-pion ratios have been performed by several collaborations, providing important constraints on transport-model descriptions of pion production in intermediate-energy HICs~\cite{SpiRIT:2021gtq,FOPI:2006ifg,HADES:2020ver,STAR:2020dav,TMEP:2022xjg}.
However, extracting reliable information on the high-density symmetry energy from pion observables remains challenging.
Noticeable discrepancies still exist between transport-model calculations and experimental data, and it remains difficult to describe all pion observables simultaneously over a broad beam-energy range~\cite{SpiRIT:2020sfn,HADES:2020ver}.
This difficulty arises because pion production depends not only on the mean-field potential, but also on the treatment of inelastic cross sections, resonance dynamics, pion optical potentials, Pauli blocking, and the initial nuclear momentum distribution.

Transport models, such as Boltzmann--Uehling--Uhlenbeck (BUU)-type models and Quantum Molecular Dynamics (QMD)-type models, have been widely used to study pion production in HICs~\cite{Li:2002qx,Ferini:2005del,Xiao:2008vm,Feng:2009am,Xie:2013np,Hong:2013yva,Song:2015hua,Godbey:2021tbt,Li:2025uku,Steinheimer:2026xeg,Xu:2013aza,Yong:2017cdl}.
In these models, the elementary $NN$ inelastic cross sections are key inputs for determining the probability of resonance production and, consequently, the final pion multiplicity.
In free space, the $NN\rightarrow N\Delta$ cross sections are usually constrained by experimental data or phenomenological parametrizations.
In dense nuclear matter, however, these reactions would be modified by the surrounding medium and the corresponding interactions \cite{Larionov:2003av,Kummer:2023hvl,Ikeno:2023cyc}.
Such in-medium effects may arise from changes in baryon effective masses, self-energies, phase space, and threshold conditions.
In our previous work~\cite{Li:2016xix}, the energy, density, and isospin dependences of the in-medium corrections to the $NN$ inelastic cross sections were calculated within the relativistic BUU (RBUU) approach~\cite{Mao:1994zza}, suggesting that such corrections may play an important role in pion production in HICs.
It has also been shown in Refs.~\cite{Godbey:2021tbt,Kummer:2023hvl} that the rapidity distributions of $\pi^{+}$ and $\pi^{-}$ measured by HADES can be well reproduced when in-medium modifications of the $NN$ inelastic cross sections are included.

Another important ingredient is the high-momentum tail (HMT) of the nucleon momentum distribution.
It is now well established that short-range correlations (SRCs), especially neutron-proton correlations, generate a sizable fraction of nucleons with momenta above the Fermi momentum in finite nuclei~\cite{Hen:2014nza,CLAS:2018xvc,Cai:2025txx,Cai:2025mrv,Fomin:2026swt,Ye:2024mls}.
Because SRC pairs are dominated by correlated $np$ pairs~\cite{Subedi:2008zz}, the HMT may also modify the isospin composition of energetic collision pairs in neutron-rich systems.
The possible influence of such high-momentum nucleons on HICs has been investigated within transport models.
Based on the isospin-dependent BUU model, it has been shown that several isospin-sensitive observables, including the difference in nucleon elliptic flows, the kinetic-energy distributions of $\pi^{-}$ and $\pi^{+}$, and the charged-pion ratio $\pi^{-}/\pi^{+}$, are sensitive to the HMT~\cite{Yong:2015gma,Zhang:2016vcc}.
In addition, Ref.~\cite{Reichert:2025egt} demonstrated that SRC-induced high-momentum components can enhance the available center-of-mass energy in individual $NN$ collisions and increase the probability of subthreshold particle production.
These studies suggest that the HMT may affect pion production by modifying the available energy in $NN$ collisions and the threshold accessibility of $NN\rightarrow N\Delta$ reactions.

Recently, it was shown in Ref.~\cite{Guo:2025xie} that both the HMT and the in-medium $NN$ cross sections can significantly affect nuclear stopping and collective flows in intermediate-energy HICs.
This indicates that the HMT and the in-medium modification of $NN$ cross sections are important ingredients in transport-model simulations.
Since pion production near threshold is closely related to the available energy in $NN$ collisions and to the probability of $NN\rightarrow N\Delta$ reactions, these two effects may also play an important role in pion observables.
However, in many transport calculations, these two factors have not been considered simultaneously, and their combined influence on pion yields, spectra, and charged-pion ratios has not been fully clarified.
Therefore, in this work, both the in-medium $NN$ inelastic cross sections calculated in our previous work~\cite{Nan:2025xvi} and the SRC-induced HMT are introduced into the Ultra-relativistic Quantum Molecular Dynamics (UrQMD) model to investigate their effects on pion production in intermediate-energy HICs.

The paper is organized as follows.
In Sec.~\ref{sec:model}, we briefly introduce the UrQMD model and describe the implementation of the HMT initialization and the in-medium $NN$ inelastic cross sections.
In Sec.~\ref{sec:results}, the calculated pion yields, spectra, ratios, and flow observables are presented, and the separate and combined effects of the two mechanisms are discussed.
Finally, the summary and outlook are given in Sec.~\ref{sec:summary}.

\section{Methodology}\label{sec:model}
\subsection{Transport model}

In this work, the UrQMD model~\cite{Bass:1998ca,Bleicher:1999xi} is used to simulate pion production in Au+Au collisions.
The density- and momentum-dependent potentials are taken in the same form as in Refs.~\cite{Li:2011zzp,Wang:2020dru,Li:2022wvu}. 
The nucleonic potential energy $U$ is calculated from the potential energy density $u$ \cite{Liu:2020jbg}, 
\begin{equation}\label{eq1}
\begin{aligned}
u=& \frac{\alpha}{2}\frac{\rho^2}{\rho_0}+\frac{\beta}{\eta+1}\frac{\rho^{\eta+1}}{\rho_0^\eta} \\
&+ \frac{g_{\rm sur}}{2\rho_0}(\nabla \rho)^2+\frac{g_{\rm sur,iso}}{2}[\nabla (\rho_n-\rho_p)]^2 \\
&+ t_{\rm md} \ln^2[1+a_{\rm md}(\textbf{p}_{i}-\textbf{p}_{j})^2]\frac{\rho}{\rho_0}\\
&+\frac{C_s}{2}(\frac{\rho}{\rho_0})^\gamma\rho\delta^2.
\end{aligned}
\end{equation}
Here, $\alpha=-211$ MeV, $\beta=113$ MeV, $\eta=1.35$, $t_{md}=3.058$ MeV,  $a_{md}=500~c^{2}/{\rm GeV}^{2}$, 
\(C_s/2=20\) MeV, \(\delta=(\rho_n-\rho_p)/(\rho_n+\rho_p)\), and \(\gamma=0.5\) are adopted, which correspond to a soft momentum-dependent EoS with an incompressibility of \(K_{0}=230~{\rm MeV}\), the slope of the symmetry energy \(L=54\) MeV. 
Both \(K_0\) and \(L\) lie in the commonly accepted range constrained by nuclear experiments and neutron-star observations~\cite{Sorensen:2023zkk,Li:2021thg,Zhang:2013wna}. 
The momentum-dependent interaction adopted here is isospin independent. Mean-field momentum dependence can modify high-momentum nucleon phase space and pion yields~\cite{Li:2025uku}, while differences between the momentum dependences of neutron and proton potentials can affect the collision terms and the $\pi^-/\pi^+$ ratio~\cite{Cozma:2021tfu,Ikeno:2023cyc}. The isovector momentum dependence and its interplay with the HMT remain sources of model uncertainty that require further investigation. 
For the \(NN\) elastic interaction, the density-, momentum-, and isospin-dependent in-medium correction factor FU3FP4 from Refs.~\cite{Li:2022wvu,Wang:2018hsw} is adopted.
This correction has been shown to play an important role in constraining the EoS below \(2\rho_0\) through comparisons between transport-model calculations and experimental data~\cite{Li:2011zzp,Russotto:2016ucm,Wang:2018hsw,Tong:2020dku}.

\subsection{High-momentum tail of the nucleon momentum distribution}
In the UrQMD model, each nucleon is represented by a Gaussian wave packet with a finite width in phase space.
The density distribution of the nuclear system is obtained by integrating the Wigner phase-space distribution over momentum space~\cite{Wang:2002ywa,Xiang:2025dxt}.
The local Fermi momentum is then calculated from the local density as \(k_F=\hbar c(3\pi^2\rho)^{1/3}\), and in the standard initialization the nucleon momenta are randomly sampled between zero and the local Fermi momentum~\cite{Bass:1998ca,Bleicher:1999xi}.

\begin{figure}[t!]
\centering
\includegraphics[width=0.45\textwidth]{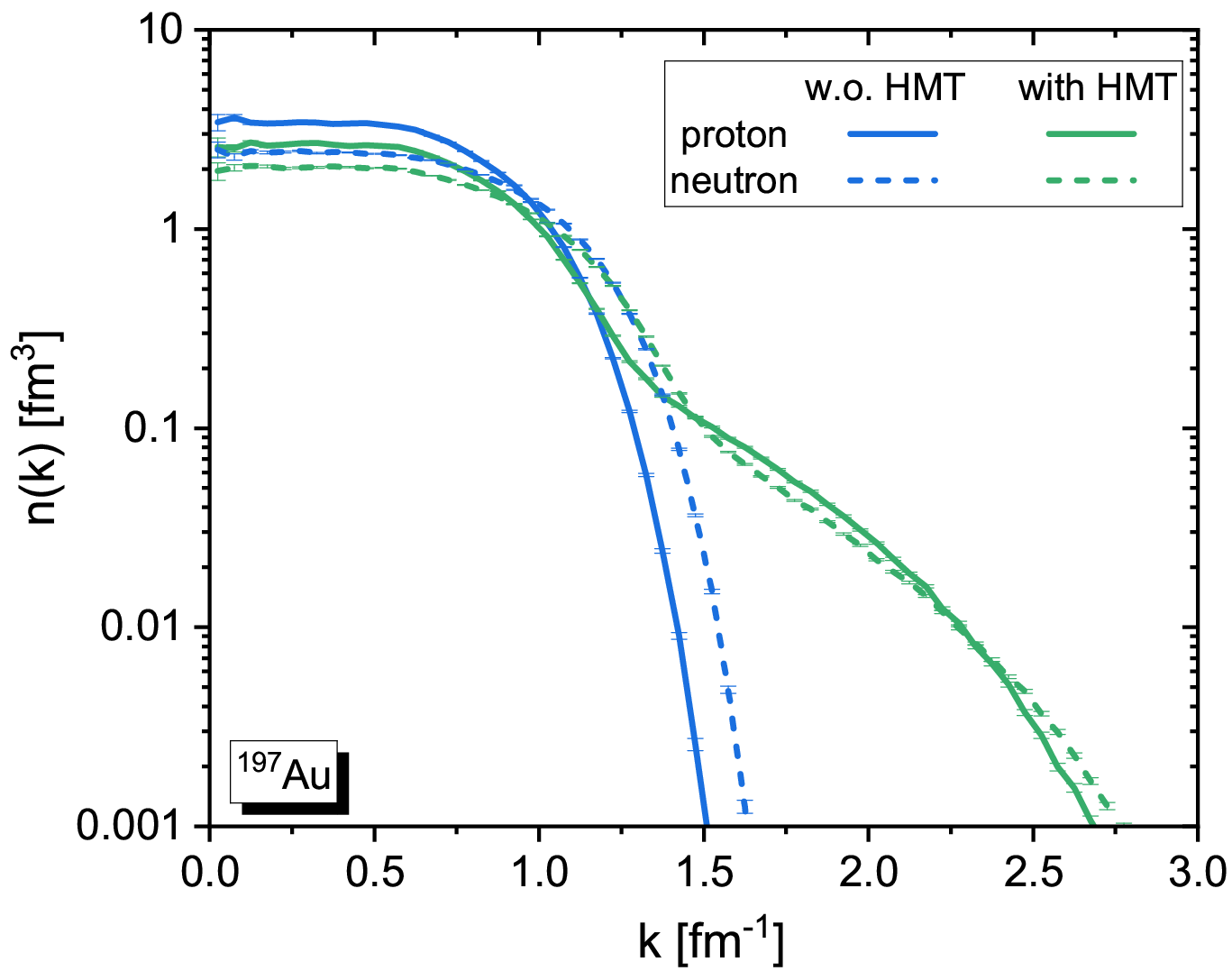}
\caption {\label{fig1_nk}(Color online) Momentum distributions \(n(k)\) of protons (solid lines) and neutrons (dashed lines) in the initialized Au nucleus with and without the HMT.
The normalization condition is \(\int_{0}^{\lambda k_F}n(k)k^2 dk=1\).}
\end{figure}

To include the SRC-induced HMT, a momentum distribution extending up to \(\lambda k_F\) is used, with the cutoff parameter set to \(\lambda=2.2\)~\cite{Yong:2015gma}.
The fraction of nucleons in the HMT is taken as \(P_{\rm HMT}=0.20\)~\cite{Zhang:2022tsw,Cai:2025mrv}. 
These parameters are adopted as phenomenological inputs for the initial nuclear momentum distribution, following earlier SRC-motivated transport calculations~\cite{Yong:2015gma,Zhang:2022tsw}. 
The isospin structure is incorporated through the preferential selection of $np$ pairs~\cite{Subedi:2008zz}.
A fixed total HMT fraction does not imply equal proton and neutron HMT fractions in a neutron-rich nucleus. 
The uncertainties associated with these parameters and their effects on HIC observables have been discussed in
Refs.~\cite{Cai:2025txx,Yong:2015gma,Yang:2018xtl,Guo:2021zcs}. 
The momentum distribution used in the initialization is
\begin{equation}
    n(k)=
    \begin{cases}
    C_{1}, & 0<k<k_{F}, \\[1ex]
    C_{2}/k^4, &
    k_{F}<k<\lambda k_{F},
    \end{cases}
\end{equation}
where \(C_1\) and \(C_2\) are normalization constants for the depleted Fermi-sea part and the HMT part, respectively, determined by the HMT fraction \(P_{\rm HMT}\) and the normalization condition of \(n(k)\). 
The initialized momentum distributions are shown in Fig.~\ref{fig1_nk}.
Compared with the standard Fermi-gas initialization, the HMT case exhibits a depletion below the Fermi momentum and a clear high-momentum component above \(k_F\), consistent with the imposed \(k^{-4}\)-type SRC tail.

\begin{table}[b]
\centering
\caption{Skyrme-type density-dependent parameters used in the calculations with and without the HMT. 
The parameters of the momentum-dependent term are kept unchanged.}
\label{tab:hmt_skyrme}
\begin{tabular}{ccccc}
\hline\hline
HMT & \(K_0\) (MeV) & \(\alpha\) (MeV) & \(\beta\) (MeV) & \(\eta\) \\
\hline
w.o.  & 230 & -211 & 113 & 1.35 \\
with  & 230 & -325 & 200 & 1.22 \\
\hline\hline
\end{tabular}
\end{table}

The introduction of the HMT will increase the average nucleon kinetic energy and change the balance between the kinetic and interaction contributions to the nuclear energy. 
The density-dependent Skyrme parameters are therefore readjusted to retain $K_0=230$~MeV, while the parameters of the momentum-dependent interaction remain unchanged.
The adopted parameter sets are listed in Tab.~\ref{tab:hmt_skyrme}. The differences between the calculations with and without the HMT therefore reflect the combined effects of the modified initial momentum distribution and the accompanying mean-field readjustment. 
A detailed assessment of their individual contributions is beyond the present work and left for future work.

\begin{figure}[t!]
\centering
\includegraphics[width=0.45\textwidth]{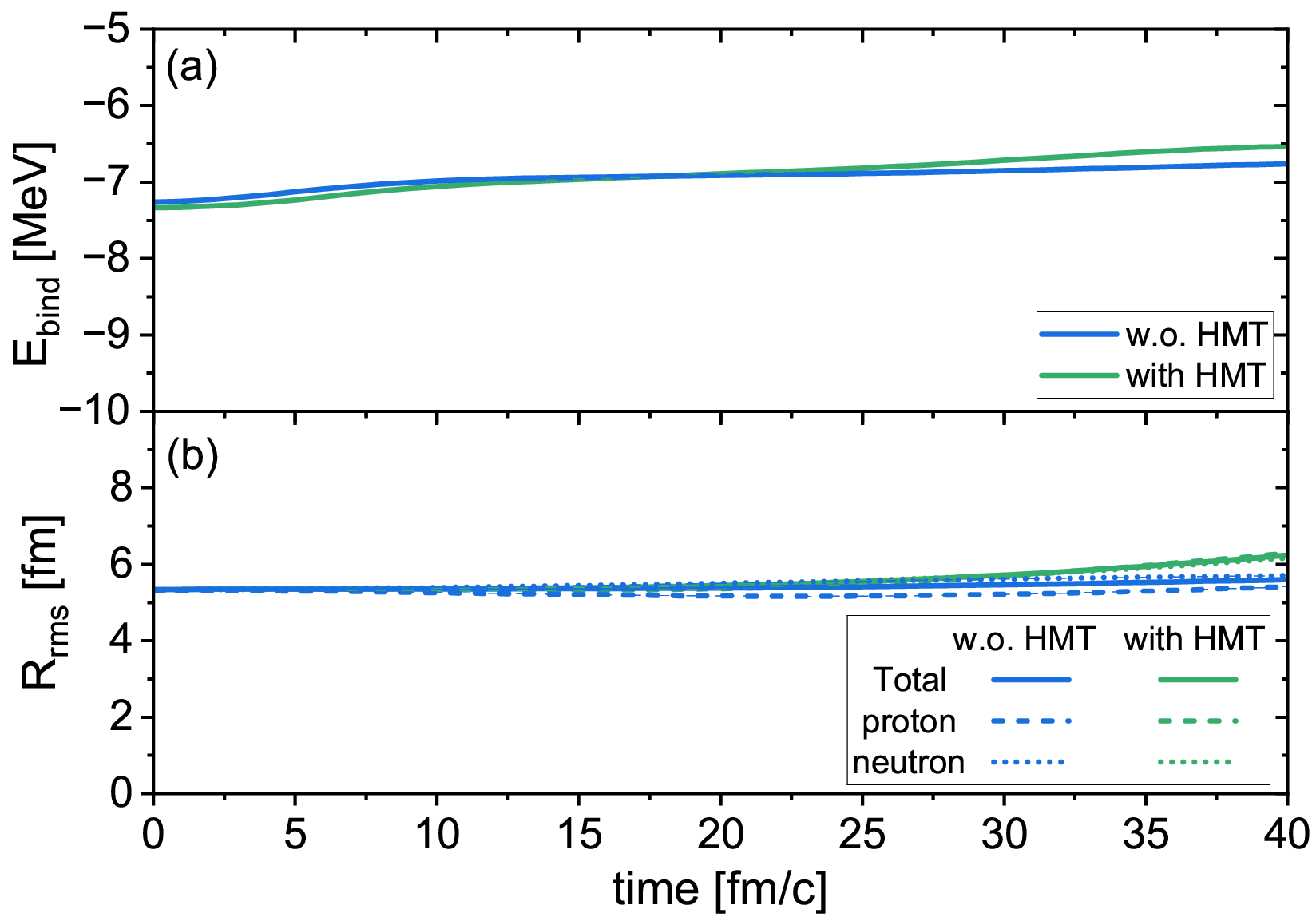}
\caption {\label{fig2_Eb_RMS}(Color online) Time evolution of the binding energy per nucleon [panel (a)] and the root-mean-square radius [panel (b)] of Au nuclei with and without the HMT.}
\end{figure}

The stability of the initialized nuclei is further examined in Fig.~\ref{fig2_Eb_RMS}.
Both the binding energy per nucleon [panel (a)] and the root-mean-square radius [panel (b)] remain approximately stable during the time evolution, indicating that the HMT initialization does not introduce significant artificial expansion or collapse.

\subsection{In-medium \(NN\rightarrow N\Delta\) cross section}

Over the past two decades, the in-medium \(NN\), \(N\Delta\) elastic, and \(NN\) inelastic cross sections have been calculated by our group within the RBUU transport theory, in which the \(\sigma\), \(\omega\), \(\rho\), and \(\delta\) meson fields are included~\cite{Li:2000sha,Li:2003vd,Li:2016xix,Li:2017pis,Nan:2024ogc,Nan:2023gwp,Nan:2025xvi}.
The in-medium \(NN\) elastic cross section has already been introduced into transport-model calculations~\cite{Li:2006ez,Li:2011zzp}, and the updated version has been used in experimental data comparisons and experimental design studies~\cite{Russotto:2011hq,Guo:2024zij}.

Based on the calculation shown in Ref.~\cite{Nan:2025xvi}, we use the following parametrization for the \(\sqrt{s}\)-dependent in-medium \(NN\rightarrow N\Delta\) cross section,
\begin{equation}\label{eqomeg}
    \sigma_{NN\rightarrow N\Delta}^{*}(\sqrt{s}) 
    =
    \begin{cases}
    3.85\exp\left[-\dfrac{(\sqrt{s}-2.121)^2}{0.0024}\right], \quad\text{for}\quad\sqrt{s} \leq 2.10 , \\[2ex]
    1.262+17.792\exp\left(0.015-\dfrac{\sqrt{s}-2.163}{0.314}\right)\\
    \quad \cdot \dfrac{1+\operatorname{erf}(z/\sqrt{2})}{2},\quad\text{for}\sqrt{s} > 2.10 ,
    \end{cases}
\end{equation}
with
\begin{equation}
    z=\dfrac{\sqrt{s}-2.163}{0.055}-0.175.
\end{equation}
This formula is obtained by fitting the cross sections calculated at \(\rho_0\) and isospin asymmetry \(\alpha=0\) with a nonlinear density-dependent coupling-constant set~\cite{Miyatsu:2022wuy,Sun:2022yor}.
This parameter set gives an EoS with \(K_0=230\) MeV. 
Although the ratio of the nucleon effective mass to the bare nucleon mass calculated with this parameter set indicates a quite strong in-medium reduction of the nucleon effective mass, it still lies in the range of other non-linear RMF parameterizations \cite{Kummer:2023hvl}. 
From calculations of the in-medium \(NN\) inelastic cross sections at different densities, a simplified density-dependent factor is introduced as
\begin{equation}
    F_\rho(u)=2.68\exp\left(-\frac{u}{1.16}\right)-0.126,\quad u=\frac{\rho}{\rho_0}.
\end{equation}
The in-medium \(NN\rightarrow N\Delta\) cross section used in the collision term is then written as
\begin{equation}\label{eqin-nnnd}
    \sigma_{NN\rightarrow N\Delta}^{\rm in}(\sqrt{s},\rho)=C_{\rm iso}\sigma_{NN\rightarrow N\Delta}^{*}(\sqrt{s}) F_\rho(u),
\end{equation}
where \(C_{\rm iso}\) is the isospin Clebsch--Gordan factor for the corresponding charge channel. 
The inverse \(N\Delta\rightarrow NN\) cross sections are obtained through detailed balance, following the standard UrQMD prescription~\cite{Bass:1998ca}.

For the reaction \(1+2\rightarrow 3+4\), the in-medium shift of the production threshold can be written as
\begin{equation}
\begin{aligned}
\Delta\sqrt{s_{\rm th}}=&\sqrt{\left(m_3+\Sigma_{s,3}+\Sigma_{0,3}+m_4+\Sigma_{s,4}+\Sigma_{0,4}\right)^2-\left|\bm{\Sigma}_3+\bm{\Sigma}_4\right|^2}\\
&-(m_3+m_4)-\left(\Sigma_{s,1}+\Sigma_{0,1}+\Sigma_{s,2}+\Sigma_{0,2}\right).
\end{aligned}
\end{equation}
The first term represents the in-medium final-state threshold, the second term is the vacuum \(N\Delta\) threshold, and the last term accounts for the scalar and vector self-energies of the two incoming nucleons.
In the present work, the spatial component of the vector self-energy is neglected for simplicity, i.e., \(\bm{\Sigma}_{B}=0\).
The effective energy used in Eq.~(\ref{eqomeg}) is then taken as \(\sqrt{s}_{\rm eff}=\sqrt{s}-\Delta\sqrt{s_{\rm th}}\).

Following the idea of treating in-medium thresholds through scalar and vector self-energies in Refs.~\cite{Ferini:2005del,Song:2015hua}, we employ a reduced RMF-like prescription for the \(NN\rightarrow N\Delta\) threshold.
For a baryon \(B\), the scalar and vector self-energies are parametrized as
\begin{equation}
    \Sigma_{s,B}=x_{s,B}\left(C_{1}+C_{2}\tau_B\delta\right)u^{C_{5}},~~
    \Sigma_{0,B}=x_{0,B}\left(C_{3}+C_{4}\tau_B\delta\right)u^{C_{5}},
\end{equation}
where \(u=\rho/\rho_0\), and \(\tau_B\) is the reduced isospin factor of baryon \(B\).
In the present convention, \(\tau_p=\tau_{\Delta^{++}}=1\), \(\tau_n=\tau_{\Delta^-}=-1\), \(\tau_{\Delta^+}=1/3\), and \(\tau_{\Delta^0}=-1/3\).
For nucleons, \(x_{s,N}=x_{0,N}=1\), while for \(\Delta\) resonances the reduced couplings are taken as \(x_{s,\Delta}=x_{0,\Delta}=2/3\), motivated by the reduced \(\Delta\) potential \(U_\Delta\simeq(2/3)U_N\) discussed in Ref.~\cite{Kummer:2023hvl}. 
In our calculations, changing $x_{s,\Delta}=x_{0,\Delta}$ from $2/3$ to $1$ modifies the absolute pion yields but has a relatively weaker effect on the $\pi^-/\pi^+$ ratio, with the latter trend qualitatively consistent with Refs.~\cite{Guo:2015tra,Ikeno:2023cyc}. 
And the in-medium suppression of pion production and its partial recovery by the HMT persist for both choices.

The values \(C_{1}=-0.350~{\rm GeV}\) and \(C_{3}=0.300~{\rm GeV}\) are adopted as reduced parameters representing the typical several-hundred-MeV scalar attraction and vector repulsion of nucleons around \(\rho_0\) in RMF and Dirac phenomenological studies~\cite{Serot:1997xg,Furnstahl:1999ff,Plohl:2006hy}.
The isovector strengths \(C_{2}=0.035~{\rm GeV}\) and \(C_{4}=-0.068~{\rm GeV}\) control the scalar and vector self-energy splittings, respectively.
Their opposite signs reflect the competing roles of the isovector scalar and vector fields in the charge-dependent threshold effect, as discussed in Refs.~\cite{Ferini:2005del,Song:2015hua}.
Several parameter sets were tested in the present work, and the above values were adopted as the baseline choice because they provide a moderate isospin-dependent threshold splitting and lead to a reasonable description of the pion observables.
The choice \(C_{5}=1\) assumes an approximately linear density dependence around \(\rho_0\).
With this setup, the charge dependence of the threshold enters through \(\tau_B\), so different channels, such as \(pp\rightarrow n\Delta^{++}\) and \(nn\rightarrow p\Delta^{-}\), can acquire different effective thresholds.

\section{Results and discussion}\label{sec:results}

In what follows, for clarity, calculations using free \(NN\) inelastic cross sections, in-medium \(NN\) inelastic cross sections, and in-medium \(NN\) inelastic cross sections combined with the HMT effect are denoted by \(S_{NN}\), \(S_{NN}^{*}\), and \(S_{NN}^{*}H\), respectively.

\subsection{Dynamical reaction process}\label{sec:31}

\begin{figure}[t]
\centering
\includegraphics[width=0.45\textwidth]{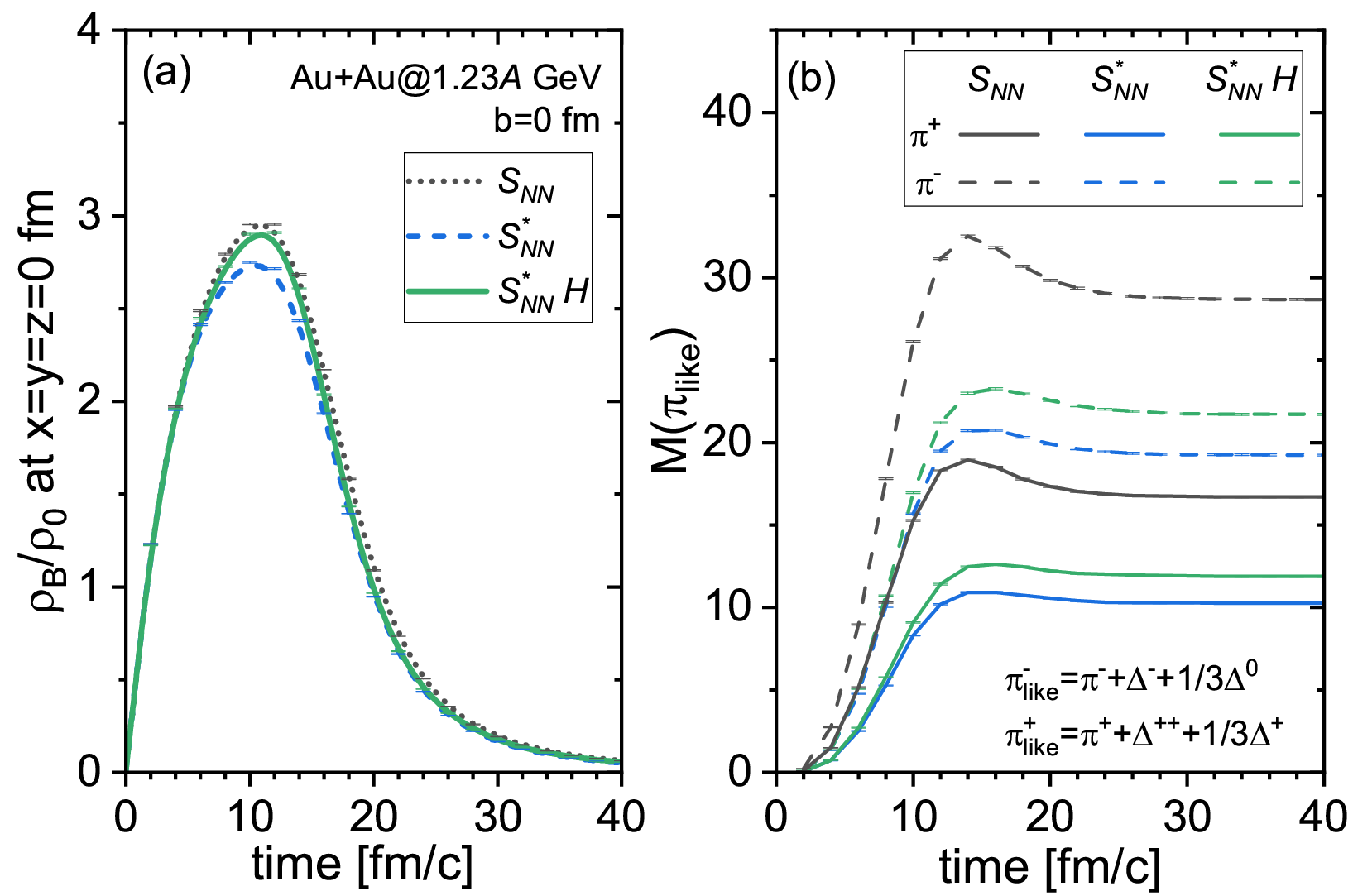}
\caption {\label{fig3_u_pilike_time}(Color online) Time evolution of the baryon density \(\rho_B/\rho_0\) and pion-like multiplicities in central (\(b=0~{\rm fm}\)) Au+Au collisions at \(E_{\rm beam}=1.23A\) GeV. The black, blue, and green lines represent the results calculated with only the in-medium \(NN\) elastic cross section, with both the in-medium \(NN\) elastic and inelastic cross sections, and with the latter setup plus the HMT, respectively. The solid and dashed lines denote \(\pi^{+}\) and \(\pi^{-}\), respectively.}
\end{figure}

\begin{figure}[t]
\centering
\includegraphics[width=0.45\textwidth]{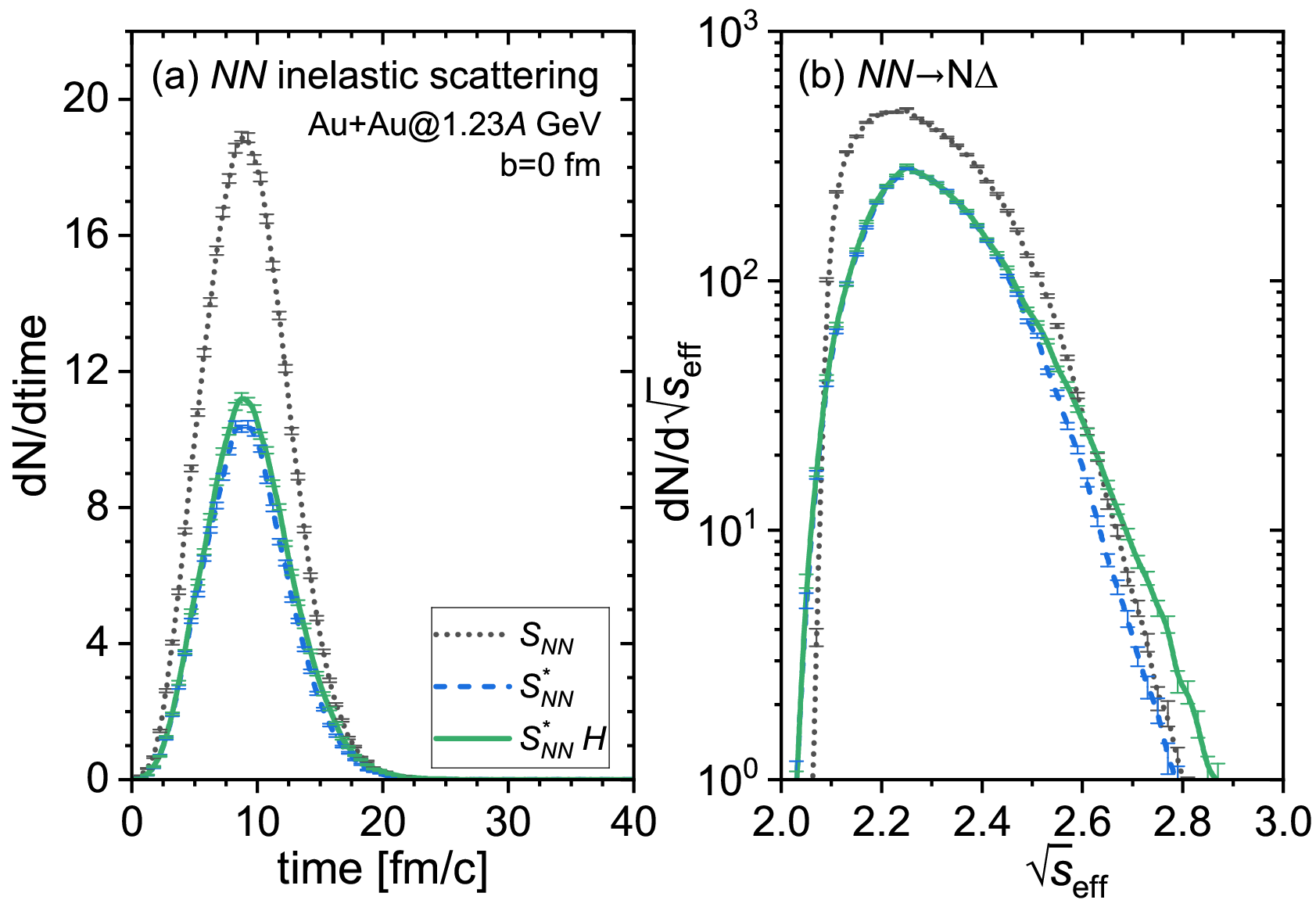}
\caption {\label{fig1_1_nn_nd_T_E}(Color online) Panel (a): time distribution of \(NN\) inelastic scatterings. Panel (b): \(\sqrt{s}_{\rm eff}\) distribution of the \(NN\to N\Delta\) channel in central (\(b=0~{\rm fm}\)) Au+Au collisions at \(E_{\rm beam}=1.23A\) GeV. The symbol sets are the same as in Figs.~\ref{fig3_u_pilike_time}~(a).}
\end{figure}

We first examine how the in-medium \(NN\rightarrow N\Delta\) cross section and the HMT affect the dynamical evolution of HICs.
Figure~\ref{fig3_u_pilike_time} shows the time evolution of the baryon density at \(x=y=z=0\) fm (left panel) and the pion-like multiplicities (right panel) in central Au+Au collisions at \(1.23A\) GeV.  
As shown in Fig.~\ref{fig3_u_pilike_time}(a), the maximum density is reached at \(t\simeq 10\)--\(12~{\rm fm}/c\) in all three cases, corresponding to the high-density stage where most \(NN\rightarrow N\Delta\) reactions occur.
For the calculation with only the in-medium \(NN\) elastic cross section, the peak density reaches about \(2.95\rho_0\).
After the in-medium \(NN\rightarrow N\Delta\) cross section is included without the HMT, the peak density is reduced to about \(2.75\rho_0\).
When the HMT is further included, the maximum density increases again to about \(2.90\rho_0\). 
In-medium reduced $NN \to N \Delta$ cross section leads to less stopping and, thus, to a lower central density. 
The HMT acts in the opposite way, increasing the rate of $NN \to N\Delta$ collisions, which results in more stopping and higher central density.
The result indicates that the HMT enhances the early collision dynamics to some extent, as also seen in Fig.~\ref{fig1_1_nn_nd_T_E}(a), and partially compensates for the reduction of compression caused by the in-medium inelastic cross section.

In Fig.~\ref{fig3_u_pilike_time}(b), the pion-like multiplicities increase rapidly during the compression stage and become nearly saturated after the system expands.
At \(t=40~{\rm fm}/c\), the in-medium \(NN\rightarrow N\Delta\) cross section reduces \(\pi^-_{\rm like}\) from 28.67 to 19.25 and \(\pi^+_{\rm like}\) from 16.70 to 10.25, corresponding to reductions of about \(33\%\) and \(39\%\), respectively.
This strong suppression reflects the reduced probability of \(N\Delta\) production in dense matter, and increases the pion-like ratio from 1.72 to 1.88.
When the HMT is introduced together with the in-medium inelastic cross section, the pion-like multiplicities are partially recovered.
At \(t=40~{\rm fm}/c\), \(\pi^-_{\rm like}\) increases from 19.25 to 21.71, while \(\pi^+_{\rm like}\) increases from 10.25 to 11.88, corresponding to enhancements of about \(13\%\) and \(16\%\), respectively.
This behavior is caused by the high-momentum components in the initial nucleon momentum distribution, which increase the available energy in individual \(NN\) collisions and make \(NN\rightarrow N\Delta\) reactions more accessible.
A similar conclusion was reached in Ref.~\cite{Reichert:2025egt}, where SRC were shown to enhance the center-of-mass energy of individual \(NN\) collisions and thereby promote subthreshold particle production in proton-nucleus collisions.
As shown in Fig.~\ref{fig1_1_nn_nd_T_E}(b), compared with the calculation without the HMT, 
the HMT broadens the \(\sqrt{s}_{\rm eff}\) distribution of inelastic $NN$ collisions and slightly enhances the high-\(\sqrt{s}_{\rm eff}\) component, leading to a modest increase in the number of successful \(NN\to N\Delta\) reactions. 
Therefore, the in-medium inelastic cross section and the HMT affect pion production in opposite directions: the former suppresses pion production, whereas the latter partly restores the pion yield by increasing the available energy in \(NN\) inelastic collisions.

\subsection{Transverse momentum distribution of charged pions}\label{sec:32}

\begin{figure}[t]
\centering
\includegraphics[width=0.45\textwidth]{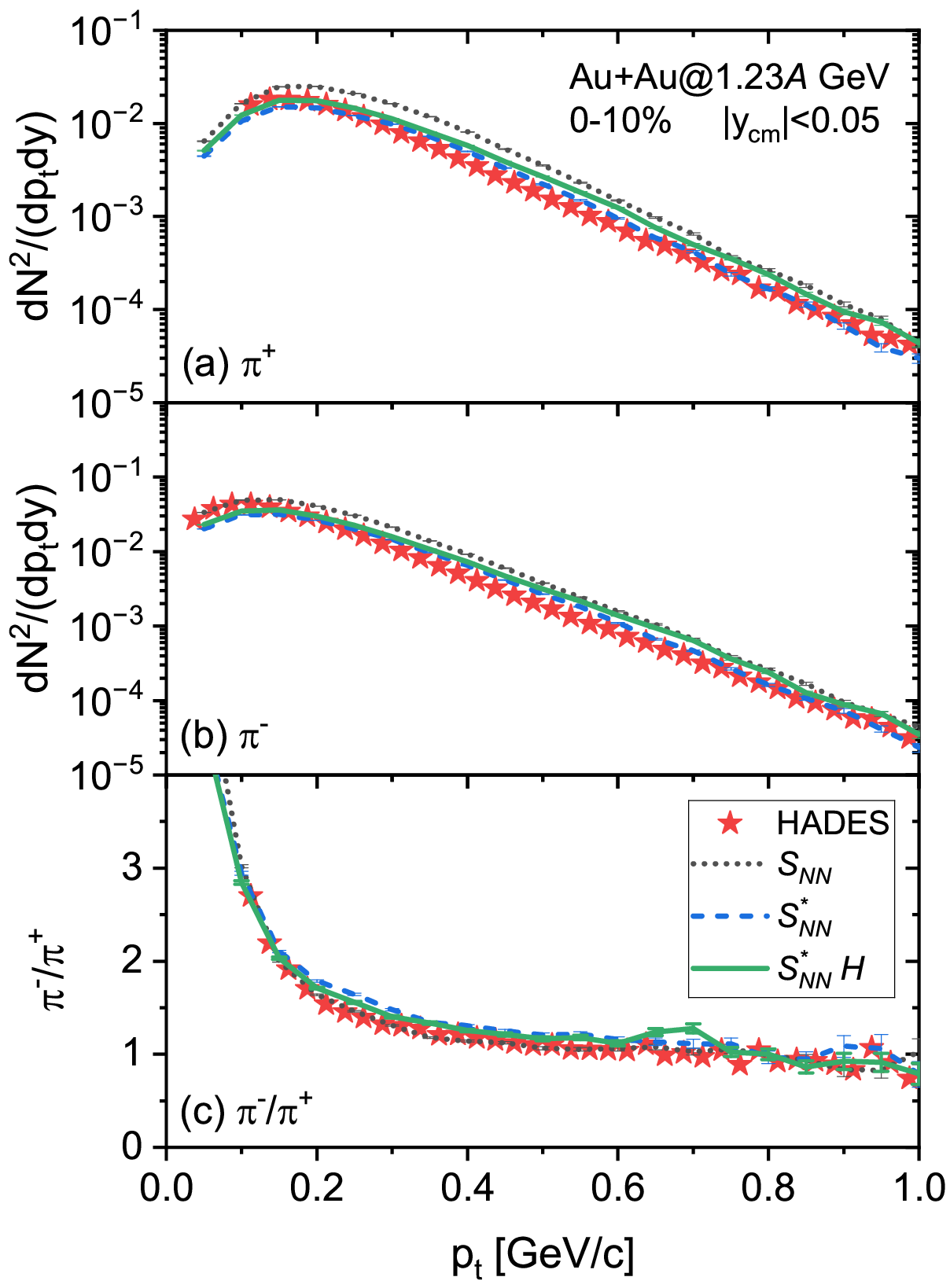}
\caption {\label{fig4_pt}(Color online) Transverse-momentum distributions of \(\pi^{+}\) (top) and \(\pi^{-}\) (middle), together with the corresponding \(\pi^{-}/\pi^{+}\) ratio (bottom), in central (\(0\)--\(10\%\)) Au+Au collisions at \(E_{\rm beam}=1.23A\) GeV. The symbols denote the HADES data~\cite{HADES:2020ver}, while the line styles and colors are the same as in Fig.~\ref{fig3_u_pilike_time}(a).}
\end{figure} 

The transverse-momentum distributions of \(\pi^{-}\) and \(\pi^{+}\), together with the \(\pi^{-}/\pi^{+}\) ratio, are shown in Fig.~\ref{fig4_pt}.
All calculations reproduce the general behavior of the HADES data~\cite{HADES:2020ver}.
Compared with the calculation that includes only the in-medium correction of the \(NN\) elastic cross section, the inclusion of the in-medium \(NN\rightarrow N\Delta\) cross section suppresses both the \(\pi^{-}\) and \(\pi^{+}\) spectra over almost the whole \(p_T\) range, bringing the results closer to the data.
After the HMT is further included, the pion spectra are partially enhanced relative to the calculation without the HMT.
The enhancement appears for both charge states and becomes more visible in the intermediate and high-\(p_T\) regions.
This behavior is consistent with the dynamical reaction process discussed in Sec.~\ref{sec:31}.

The \(p_T\)-dependent \(\pi^{-}/\pi^{+}\) ratio is less sensitive to the overall suppression or enhancement of pion production, because part of the common change in \(\pi^{-}\) and \(\pi^{+}\) cancels in the ratio. 
Nevertheless, the in-medium \(NN\rightarrow N\Delta\) cross section slightly increases the \(\pi^{-}/\pi^{+}\) ratio in the intermediate-\(p_T\) region, indicating a charge-dependent modification of pion production. 
In our other work, it is shown that this \(p_T\)-dependent ratio is strongly affected by the Coulomb interaction \cite{lpcHBT}. 
With the inclusion of the HMT, the ratio is slightly reduced relative to the calculation without the HMT, because the recovery of the \(\pi^{+}\) spectrum is somewhat stronger than that of the \(\pi^{-}\) spectrum in this region. 
The kinetic-energy distributions of charged pions have also been calculated and show the same systematic behavior as the transverse-momentum spectra. They are therefore not shown here for brevity.

\subsection{Rapidity distribution of charged pions}\label{sec:33}
\begin{figure}[t]
\centering
\includegraphics[width=0.48\textwidth]{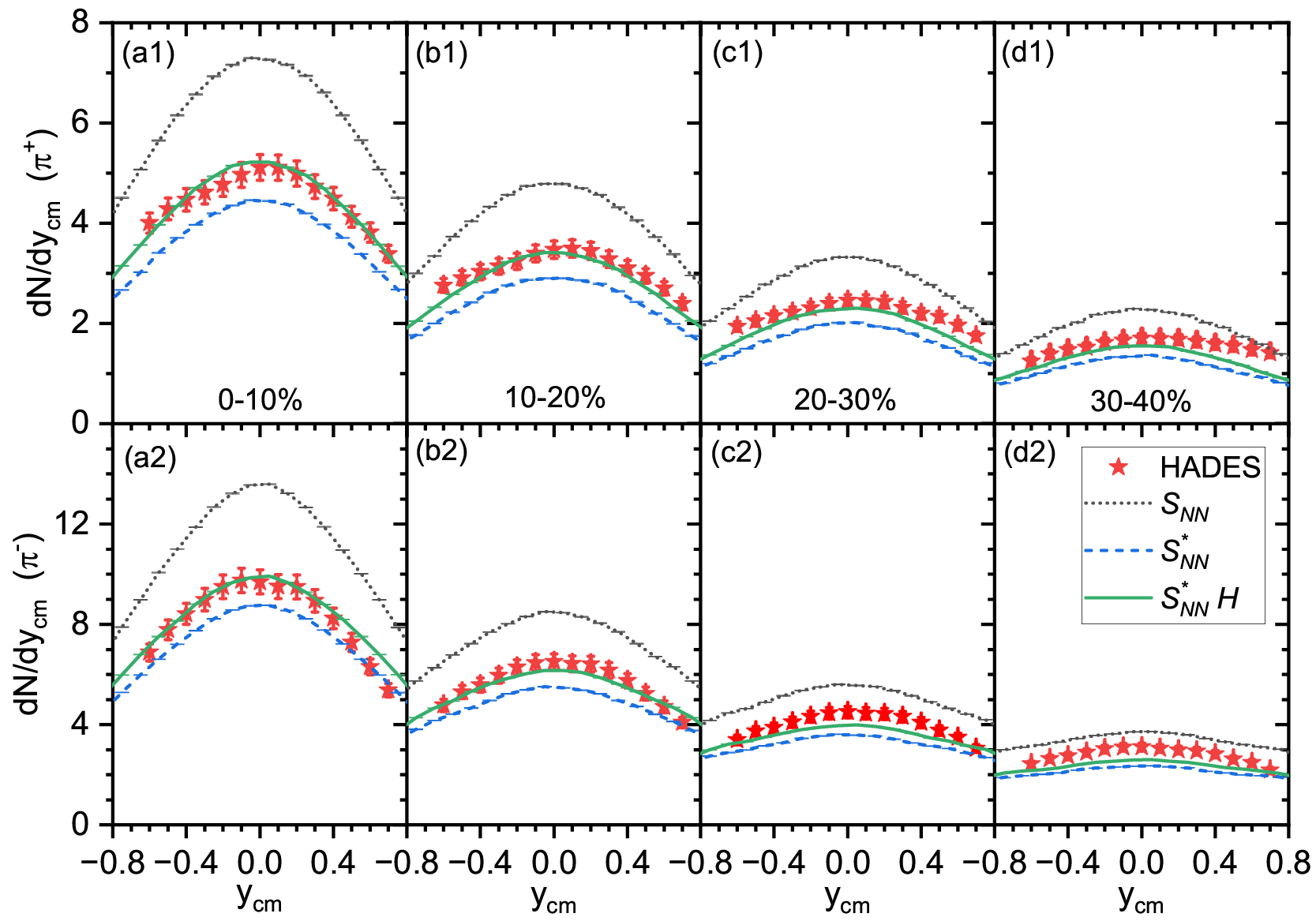}
\caption {\label{fig5_dndy_ycm}(Color online) Rapidity distributions of \(\pi^+\) (top) and \(\pi^-\) (bottom) in Au+Au collisions at \(E_{\rm beam}=1.23A\) GeV for four centrality classes. The symbols denote the HADES data~\cite{HADES:2020ver}, while the lines represent UrQMD calculations with different treatments of the in-medium effects and the HMT.}
\end{figure} 

The HADES rapidity distributions of charged pions were shown to be substantially overestimated by several transport models~\cite{HADES:2020ver}.
Therefore, we further calculate the rapidity distributions of \(\pi^-\) and \(\pi^+\) in Au+Au collisions at \(1.23A\) GeV for four centrality classes and compare them with the HADES data.
As shown in Fig.~\ref{fig5_dndy_ycm}, the calculation with only the in-medium correction of \(NN\) elastic cross sections clearly overestimates the pion rapidity distributions for both charge states.
The discrepancy appears over the whole measured rapidity region, indicating that the treatment of the \(NN\rightarrow N\Delta\) inelastic channel is important for pion production near threshold~\cite{Song:2015hua,Godbey:2021tbt,Kummer:2023hvl,Larionov:2003av}.
After the in-medium \(NN\rightarrow N\Delta\) cross section is introduced, the pion rapidity distributions are strongly suppressed.
However, this suppression tends to make the calculated distributions lower than the HADES data, especially outside the most central collisions.
When the HMT is further included, the pion rapidity distributions are enhanced again and move closer to the experimental data.
This effect is caused by the high-momentum components of nucleons in the initial state, which increase the available energy in individual \(NN\) collisions and make \(NN\rightarrow N\Delta\) reactions more accessible, as discussed in Sec.~\ref{sec:31} and shown in Figs.~\ref{fig3_u_pilike_time} and~\ref{fig1_1_nn_nd_T_E}.

It should be pointed out that the published HADES pion yields may require an upward revision of approximately 20\%, due to an underestimation of the presence of $\delta$-electrons preventing the detection of all pions ~\cite{hades,Arnaud}.
Since updated rapidity distributions of charged pions are not yet available, we retain the original published data in Fig.~\ref{fig5_dndy_ycm}. 
A uniform 20\% upward rescaling would place the data points between the predictions for $S_{NN}$ and $S_{NN}^{*}H$. 
And the qualitative conclusion of the present calculations remains unchanged: the in-medium correction of the $NN$ inelastic cross section suppresses pion production, whereas the HMT partly restores the pion yield.  
These results demonstrate that pion production at the investigated energy is sensitive to both the in-medium inelastic cross sections and the SRC-induced high-momentum components \cite{Yong:2015wha}.
A more quantitative discussion of the total pion yields, the charged-pion ratios, and their beam-energy and centrality dependence is given below.
\subsection{Centrality- and beam energy-dependence of charged-pion yields}\label{sec:34}
\begin{figure}[t]
\centering
\includegraphics[width=0.48\textwidth]{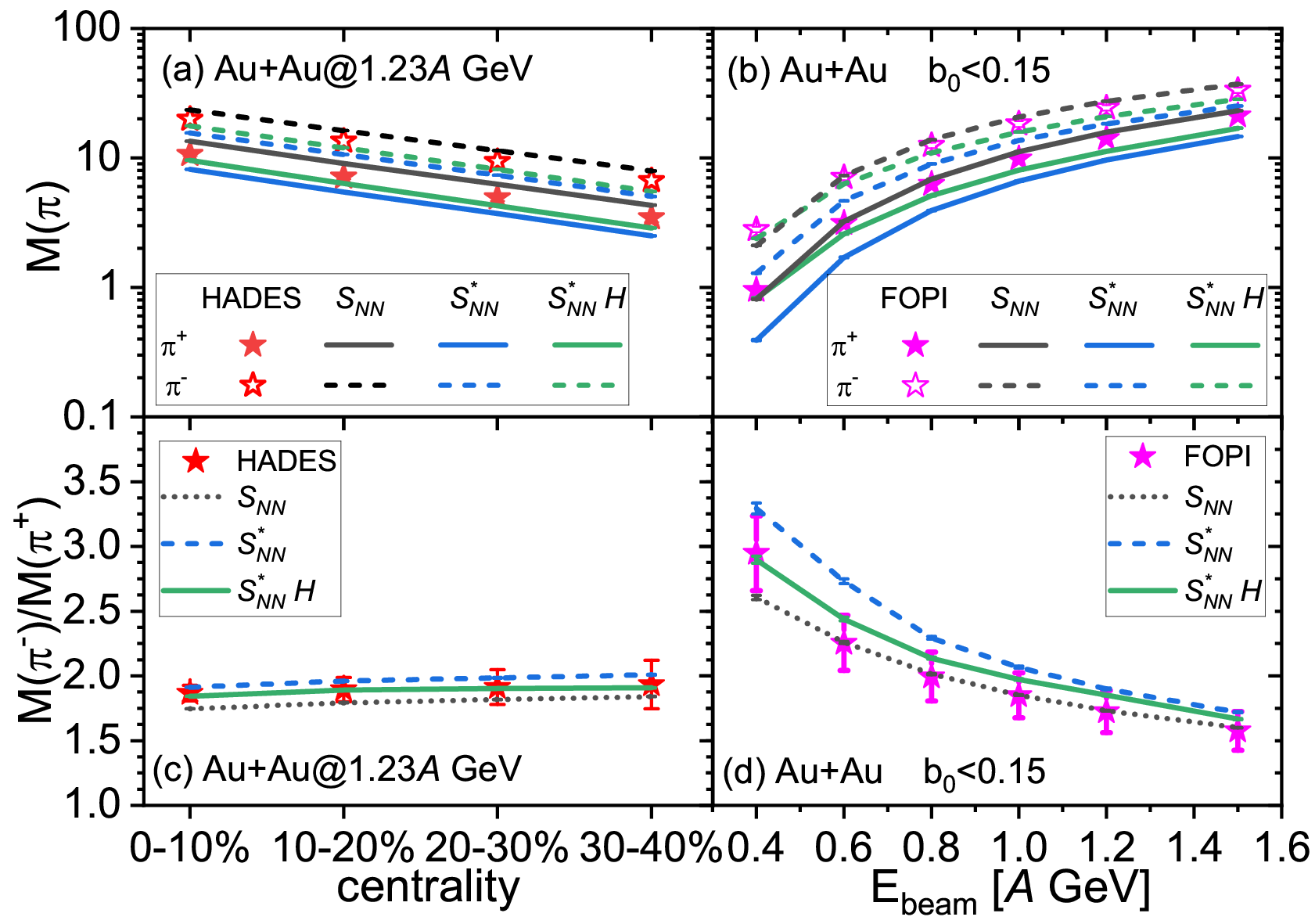}
\caption {\label{fig6_cent_energy_pion}(Color online) Centrality and beam-energy dependences of pion multiplicities and the charged-pion ratio in Au+Au collisions. Panels (a) and (c) show the centrality dependence of the \(\pi^{+}\), \(\pi^{-}\) multiplicities and the \(\pi^{-}/\pi^{+}\) ratio at \(1.23A\) GeV, respectively. Panels (b) and (d) show the beam-energy dependence of the corresponding observables for central collisions with \(b_{0}<0.15\). The red and magenta star symbols denote the HADES preliminary data~\cite{hades} and FOPI data~\cite{FOPI:2006ifg}.}
\end{figure} 

The centrality and beam-energy dependences of the charged-pion multiplicities and the \(\pi^{-}/\pi^{+}\) ratio are shown in Fig.~\ref{fig6_cent_energy_pion}.
The corrected HADES data at \(1.23A\) GeV cover four centrality classes from \(0\)--\(40\%\), while the FOPI data provide the beam-energy dependence for central Au+Au collisions.
These observables offer an integrated test of the pion-production mechanism discussed in Secs.~\ref{sec:31} and~\ref{sec:32}.
In particular, they are sensitive to the total number of successful \(NN\rightarrow N\Delta\) reactions, the subsequent \(\Delta\leftrightarrow N\pi\) dynamics, and the charge dependence of the production channels.

As shown in Fig.~\ref{fig6_cent_energy_pion}(a), both \(\pi^{-}\) and \(\pi^{+}\) multiplicities decrease from central to peripheral collisions.
This behavior mainly reflects the reduction of participant matter, density, and the number of inelastic \(NN\) collisions in more peripheral events~\cite{FOPI:2006ifg,HADES:2020ver}.
The calculation using only the in-medium \(NN\) elastic cross section overestimates the charged-pion yields.
After the in-medium \(NN\rightarrow N\Delta\) cross section is introduced, the pion multiplicities are strongly reduced because the probability of producing \(\Delta\) resonances in dense matter is suppressed.
When the HMT is further included, the pion yields are partially recovered and become closer to the HADES data over the whole centrality range.

In Fig.~\ref{fig6_cent_energy_pion}(b), the charged-pion multiplicities increase rapidly with beam energy, as expected for pion production near threshold, because the available phase space for \(NN\rightarrow N\Delta\) and the subsequent \(\Delta\rightarrow N\pi\) decay grows strongly with energy~\cite{FOPI:2006ifg,Hong:2013yva}.
The calculation with only the in-medium \(NN\) elastic cross section gives the largest pion yields, while the in-medium \(NN\rightarrow N\Delta\) cross section suppresses the yields at all beam energies.
The HMT compensates part of this suppression, especially at lower beam energies, where pion production is more sensitive to the high-momentum components of the nucleon momentum distribution.
A similar beam-energy dependence of the HMT effect on nuclear stopping was reported in Ref.~\cite{Guo:2025xie}.
This behavior supports the conclusion drawn from Secs.~\ref{sec:31} and~\ref{sec:32} that the in-medium inelastic cross section and the HMT affect pion production in opposite directions.

The corresponding centrality and beam-energy dependences of the \(\pi^{-}/\pi^{+}\) ratio are shown in Figs.~\ref{fig6_cent_energy_pion}(c) and~(d).
The calculation using only the in-medium \(NN\) elastic cross section yields a smaller ratio, while the in-medium \(NN\rightarrow N\Delta\) cross section increases the ratio.
This indicates that the in-medium modification of the inelastic channel does not suppress \(\pi^{-}\) and \(\pi^{+}\) production in exactly the same way.
The HMT lowers the ratio relative to the calculation without the HMT because the recovery of \(\pi^{+}\) production is slightly stronger than that of \(\pi^{-}\) production, as shown in Fig.~\ref{fig3_u_pilike_time}(b).
As a result, the calculation including both the in-medium inelastic cross section and the HMT gives a reasonable description of the centrality and beam-energy dependences of the charged-pion ratio data.

Overall, existing data seem to support the need to consider both ingredients, although additional more detailed theoretical studies would be probably needed after the tension between the HADES and FOPI data will be resolved.
The in-medium \(NN\rightarrow N\Delta\) cross section is required to suppress the overproduction of pions, while the HMT restores part of the pion yield and improves the charged-pion ratio.
This conclusion is consistent with recent transport-model studies showing that the description of pion observables at SIS energies depends sensitively on the treatment of in-medium cross sections, mean fields, and resonance dynamics~\cite{Godbey:2021tbt,Kummer:2023hvl,Li:2025uku,Steinheimer:2026xeg}.

\subsection{Collective flows of protons and charged pions}\label{sec:35}
\begin{figure}[t]
\centering
\includegraphics[width=0.48\textwidth]{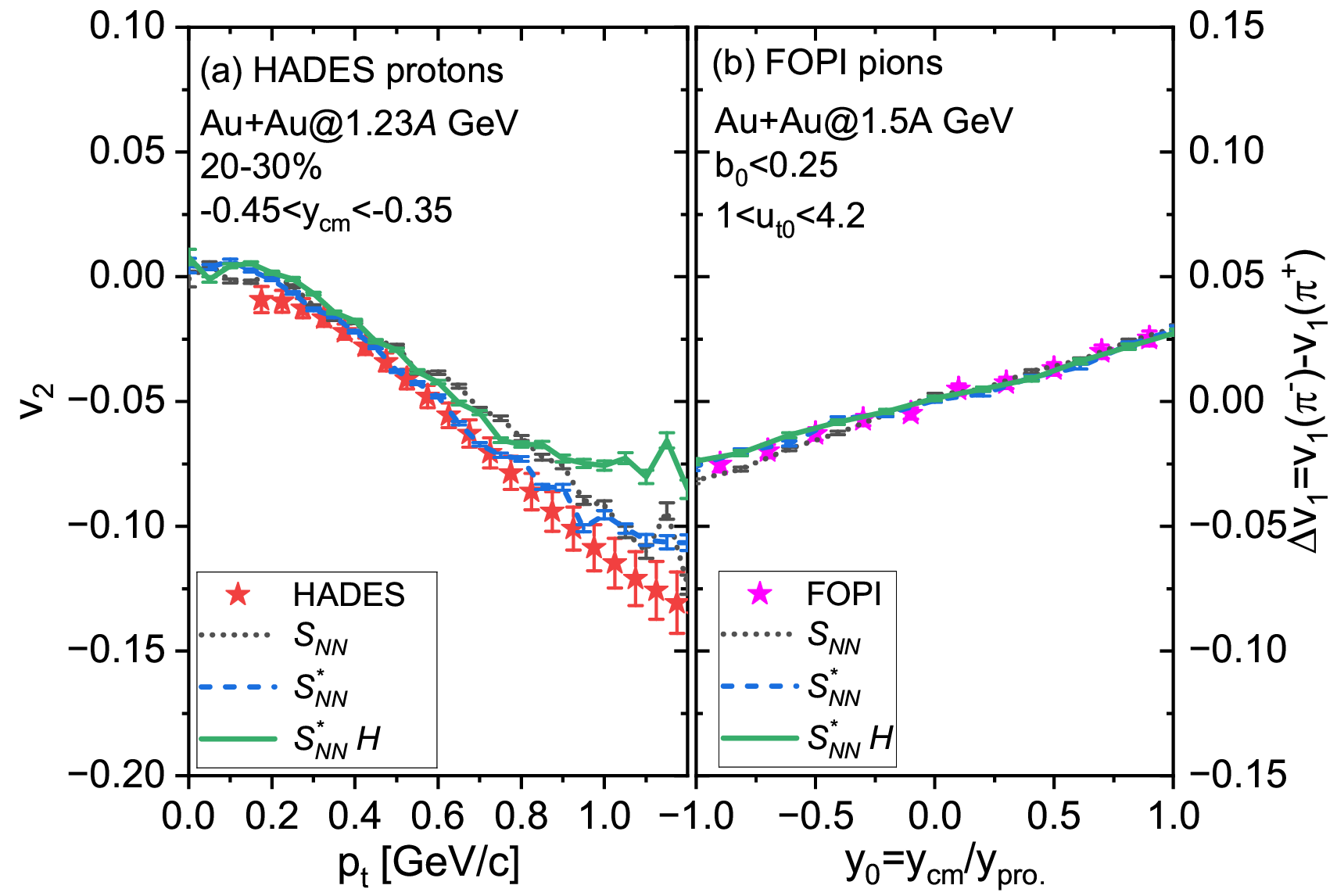}
\caption {\label{fig7_v2pt_v1_y}(Color online) Left panel: proton elliptic flow \(v_2\) as a function of transverse momentum \(p_T\) in Au+Au collisions at \(1.23A\) GeV for \(20\)--\(30\%\) centrality and \(-0.45<|y_{cm}|<-0.35\), compared with the HADES data~\cite{HADES:2022osk}. Right panel: the difference between the directed flow of \(\pi^-\) and \(\pi^+\), \(\Delta v_1=v_1(\pi^-)-v_1(\pi^+)\), as a function of reduced rapidity \(y_0\) in central Au+Au collisions at \(1.5A\) GeV, compared with the FOPI data~\cite{FOPI:2006ifg}. The symbol sets are the same as in Figs.~\ref{fig6_cent_energy_pion}(c) and~(d).}
\end{figure} 

To further test the influence of the in-medium effects and the HMT on intermediate-energy HICs, the transverse-momentum- and rapidity-dependent collective flows are shown in Fig.~\ref{fig7_v2pt_v1_y}.
The left panel presents the proton elliptic flow \(v_2\) as a function of transverse momentum in Au+Au collisions at \(1.23A\) GeV. 
The negative elliptic flow is the typical squeeze-out signal at SIS energies and is caused by the competition between the pressure developed in the compressed participant matter and the shadowing of the spectator matter~\cite{Wang:2024ktk,Reichert:2024ayg}. 
In the low-\(p_T\) region, the three calculations give very similar results.
This indicates that the low-\(p_T\) proton elliptic flow in the investigated rapidity window is not very sensitive to the in-medium correction of the \(NN\rightarrow N\Delta\) cross section or to the HMT.
This part of the spectrum mainly reflects the bulk compression-expansion dynamics and is therefore expected to be more sensitive to the nuclear EoS and the mean-field potential.
For \(p_T\gtrsim 0.4~{\rm GeV}/c\), visible differences among the three calculations begin to appear.
The in-medium \(NN\rightarrow N\Delta\) cross section slightly changes the proton \(v_2\).
After the HMT is further included, the calculated proton \(v_2\) becomes larger, especially for \(p_T\gtrsim 1.0~{\rm GeV}/c\). 
The effects of these two factors on the directed flow of protons are similar to those shown here,  therefore, they are not shown here. 
This behavior can be understood from the fact that the HMT introduces high-momentum nucleons in the initial nuclei.
These nucleons can participate in early \(NN\rightarrow NN\) and \(NN\rightarrow N\Delta\) collisions with larger relative momenta, and high-\(p_T\) protons are more sensitive to such early-stage collision dynamics. 
Similar sensitivities of stopping and collective flows to the HMT and in-medium \(NN\) cross sections have also been reported in Ref.~\cite{Guo:2025xie}.

The right panel shows the rapidity dependence of \(\Delta v_1=v_1(\pi^-)-v_1(\pi^+)\) in Au+Au collisions at \(1.5A\) GeV, compared with the FOPI data~\cite{FOPI:2006ifg}.
In the present study, the pion directed flow difference shows little sensitivity to the in-medium correction of the \(NN\rightarrow N\Delta\) cross section or to the HMT. The three calculations are close to one another within the present accuracy.
This is an interesting feature because it suggests that \(\Delta v_1\) may retain its sensitivity to the isovector mean field and may therefore serve as a useful observable for constraining the high-density symmetry energy.
However, this conclusion should be taken with caution. 
The in-medium \(NN\rightarrow N\Delta\) cross section used in Eq.~(\ref{eqin-nnnd}) does not contain an explicit isospin dependence beyond the Clebsch--Gordan factor and the threshold treatment.
Therefore, a more complete treatment with explicitly isospin- and density-dependent in-medium inelastic cross sections is needed in future work before this observable can be used for quantitative constraints on the symmetry energy.

\section{Summary and outlook}\label{sec:summary}

In this work, the in-medium \(NN\) inelastic cross sections and the SRC-induced HMT of the nucleon momentum distribution are introduced into the UrQMD model to investigate pion production in intermediate-energy HICs.
The in-medium correction of the \(NN\rightarrow N\Delta\) cross section suppresses pion production by reducing the probability of resonance production in dense matter, whereas the HMT partly restores the pion yield by increasing the high-momentum components of nucleons and the available energy in early \(NN\) collisions.
With the simultaneous inclusion of these two effects, the rapidity distributions, transverse-momentum spectra, centrality dependence, beam-energy dependence, and charged-pion ratios can be better described.
These results indicate that both the in-medium correction of \(NN\) inelastic cross sections and the initial high-momentum components are important for a consistent description of pion production near the threshold. 
In addition, the proton elliptic flow at low-\(p_T\) is not very sensitive to the HMT or to the in-medium \(NN\rightarrow N\Delta\) cross section, whereas visible effects appear at higher \(p_T\).
And the directed flow difference between charged-pions, \(\Delta v_1=v_1(\pi^-)-v_1(\pi^+)\), shows only weak sensitivity to these two effects. 
This suggests that \(\Delta v_1\) may still be useful for constraining the isovector part of the nuclear mean field, although a more complete treatment of charge-dependent in-medium inelastic cross sections is required.

The present results highlight the competing effects of in-medium $NN$ inelastic cross sections and the HMT on pion production at GeV energies. Several issues should be addressed in future studies. 
First, the HMT initialization should be better constrained by nuclear-structure information, including the momentum distribution, SRC fraction, binding energy, and rms radius. Its parameter and isospin dependences should be quantified, and its direct effects should be disentangled from those of the accompanying mean-field readjustment. 
Second, the density and isospin dependence of the in-medium \(NN\) inelastic cross section would be improved. The cross sections calculated over a broader density, energy, and isospin range will be adopted. Third, the single-particle potentials, in-medium cross sections, and reaction thresholds should be treated consistently within a common relativistic framework, with particular attention to the isovector momentum dependence and its interplay with the HMT. Finally, the combined effects of the HMT, the nuclear EoS, and the in-medium elastic and inelastic cross sections on observables, such as collective flows and femtoscopic correlations, could be investigated within a unified framework, which is essential for extracting reliable constraints on the high-density nuclear EoS.
\section*{Acknowledgements}
We thank Dr. A.~Le F{\`e}vre and Dr. M.~Lorenz for helpful discussions and valuable comments on the experimental data.
This work is supported in part by the National Natural Science Foundation of China under Grant 12505143, 12335008, and 12675166, the National Key Research and Development Program of China under Grant No. 2023YFA1606402, and the Zhejiang Provincial Natural Science Foundation of China under Grant No. LQN25A050003.
The authors are grateful to the C3S2 computing center at Huzhou Normal University for computational support.
P.~C.~Li gratefully acknowledges financial support from the China Scholarship Council under Grant No.~202608330358.


\begin{thebibliography}{00}
\expandafter\ifx\csname url\endcsname\relax
  \def\url#1{\texttt{#1}}\fi
\expandafter\ifx\csname urlprefix\endcsname\relax\def\urlprefix{URL
}\fi \expandafter\ifx\csname href\endcsname\relax
  \def\href#1#2{#2} \def\path#1{#1}\fi

\bibitem{Huth:2021bsp}
S.~Huth, P.~T.~H.~Pang, I.~Tews, \textit{et al.}
Nature \textbf{606} (2022), 276-280
doi:10.1038/s41586-022-04750-w

\bibitem{Tsang:2023vhh}
C.~Y.~Tsang, M.~B.~Tsang, W.~G.~Lynch, R.~Kumar and C.~J.~Horowitz,
Nature Astron. \textbf{8} (2024), 328-336
doi:10.1038/s41550-023-02161-z

\bibitem{Danielewicz:2002pu}
P.~Danielewicz, R.~Lacey and W.~G.~Lynch,
Science \textbf{298} (2002), 1592-1596
doi:10.1126/science.1078070

\bibitem{Sorensen:2023zkk}
A.~Sorensen, K.~Agarwal, K.~W.~Brown, \textit{et al.}
Prog. Part. Nucl. Phys. \textbf{134} (2024), 104080
doi:10.1016/j.ppnp.2023.104080

\bibitem{Niksic:2011sg}
T.~Niksic, D.~Vretenar and P.~Ring,
Prog. Part. Nucl. Phys. \textbf{66} (2011), 519-548
doi:10.1016/j.ppnp.2011.01.055

\bibitem{Dietrich:2020efo}
T.~Dietrich, M.~W.~Coughlin, P.~T.~H.~Pang, M.~Bulla, J.~Heinzel, L.~Issa, I.~Tews and S.~Antier,
Science \textbf{370} (2020), 1450-1453
doi:10.1126/science.abb4317

\bibitem{Harris:1984up}
J.~W.~Harris, R.~Stock, R.~Bock, \textit{et al.}
Phys. Lett. B \textbf{153} (1985), 377-381
doi:10.1016/0370-2693(85)90476-9

\bibitem{Li:2005gfa}
Q.~F.~Li, Z.~X.~Li, S.~Soff, M.~Bleicher and H.~Stoecker,
J. Phys. G \textbf{32} (2006) 151-164
doi:10.1088/0954-3899/32/2/007

\bibitem{Li:2008gp}
B.~A.~Li, L.~W.~Chen and C.~M.~Ko,
Phys. Rept. \textbf{464} (2008), 113-281
doi:10.1016/j.physrep.2008.04.005

\bibitem{SpiRIT:2021gtq}
J.~Estee \textit{et al.} [S$\pi$RIT Collaboration],
Phys. Rev. Lett. \textbf{126} (2021), 162701
doi:10.1103/PhysRevLett.126.162701

\bibitem{FOPI:2006ifg}
W.~Reisdorf \textit{et al.} [FOPI Collaboration],
Nucl. Phys. A \textbf{781} (2007), 459-508
doi:10.1016/j.nuclphysa.2006.10.085

\bibitem{HADES:2020ver}
J.~Adamczewski-Musch \textit{et al.} [HADES Collaboration],
Eur. Phys. J. A \textbf{56} (2020), 259
doi:10.1140/epja/s10050-020-00237-2

\bibitem{STAR:2020dav}
J.~Adam \textit{et al.} [STAR Collaboration],
Phys. Rev. C \textbf{103} (2021), 034908
doi:10.1103/PhysRevC.103.034908

\bibitem{TMEP:2022xjg}
H.~Wolter \textit{et al.} [TMEP Collaboration],
Prog. Part. Nucl. Phys. \textbf{125} (2022), 103962
doi:10.1016/j.ppnp.2022.103962

\bibitem{SpiRIT:2020sfn}
G.~Jhang \textit{et al.} [S$\pi$RIT and TMEP Collaborations],
Phys. Lett. B \textbf{813} (2021), 136016
doi:10.1016/j.physletb.2020.136016

\bibitem{Li:2002qx}
B.~A.~Li,
Phys. Rev. Lett. \textbf{88} (2002), 192701
doi:10.1103/PhysRevLett.88.192701

\bibitem{Ferini:2005del}
G.~Ferini, M.~Colonna, T.~Gaitanos and M.~Di Toro,
Nucl. Phys. A \textbf{762} (2005), 147-166
doi:10.1016/j.nuclphysa.2005.08.007

\bibitem{Xiao:2008vm}
Z.~G. Xiao, B.~A.~Li, L.~W.~Chen, G.~C.~Yong and M.~Zhang,
Phys. Rev. Lett. \textbf{102} (2009), 062502
doi:10.1103/PhysRevLett.102.062502

\bibitem{Feng:2009am}
Z.~Q.~Feng and G.~M.~Jin,
Phys. Lett. B \textbf{683} (2010), 140-144
doi:10.1016/j.physletb.2009.12.006

\bibitem{Yong:2017cdl}
G.~C.~Yong,
Phys. Rev. C \textbf{96} (2017), 044605
doi:10.1103/PhysRevC.96.044605

\bibitem{Xu:2013aza}
J.~Xu, L.~W.~Chen, C.~M.~Ko, B.~A.~Li and Y.~G.~Ma,
Phys. Rev. C \textbf{87} (2013), 067601
doi:10.1103/PhysRevC.87.067601

\bibitem{Xie:2013np}
W.~J.~Xie, J.~Su, L.~Zhu and F.~S.~Zhang,
Phys. Lett. B \textbf{718} (2013), 1510-1514
doi:10.1016/j.physletb.2012.12.021

\bibitem{Hong:2013yva}
J.~Hong and P.~Danielewicz,
Phys. Rev. C \textbf{90} (2014), 024605
doi:10.1103/PhysRevC.90.024605

\bibitem{Song:2015hua}
T.~Song and C.~M.~Ko,
Phys. Rev. C \textbf{91} (2015), 014901
doi:10.1103/PhysRevC.91.014901

\bibitem{Godbey:2021tbt}
K.~Godbey, Z.~Zhang, J.~W.~Holt and C.~M.~Ko,
Phys. Lett. B \textbf{829} (2022), 137134
doi:10.1016/j.physletb.2022.137134

\bibitem{Li:2025uku}
X.~Li, S.~P.~Wang, Z.~Zhang, R.~Wang, J.~Pu, C.~W.~Ma and L.~W.~Chen,
Phys. Lett. B \textbf{872} (2026), 140114
doi:10.1016/j.physletb.2025.140114

\bibitem{Steinheimer:2026xeg}
J.~Steinheimer and M.~Bleicher,
Phys. Rev. C \textbf{114} (2026), 044904.
doi:10.1103/6j14-cm4m

\bibitem{Larionov:2003av}
A.~B.~Larionov and U.~Mosel,
Nucl. Phys. A \textbf{728} (2003), 135-164
doi:10.1016/j.nuclphysa.2003.08.005

\bibitem{Kummer:2023hvl}
C.~Kummer, K.~Gallmeister and L.~von Smekal,
Phys. Rev. C \textbf{109} (2024), 054901
doi:10.1103/PhysRevC.109.054901

\bibitem{Ikeno:2023cyc}
N.~Ikeno and A.~Ono,
Phys. Rev. C \textbf{108} (2023), 044601
doi:10.1103/PhysRevC.108.044601

\bibitem{Li:2016xix}
Q.~F. Li and Z.~X. Li,
Phys. Lett. B \textbf{773} (2017), 557-562
doi:10.1016/j.physletb.2017.09.013

\bibitem{Mao:1994zza}
G.~J. Mao, Z.~X. Li, Y.~Z. Zhuo, Y.~L. Han and Z.~Yu,
Phys. Rev. C \textbf{49} (1994), 3137-3146
doi:10.1103/PhysRevC.49.3137

\bibitem{Hen:2014nza}
O.~Hen, M.~Sargsian, L.~B.~Weinstein, \textit{et al.}
Science \textbf{346} (2014), 614-617
doi:10.1126/science.1256785

\bibitem{CLAS:2018xvc}
M.~Duer \textit{et al.} [CLAS Collaboration],
Phys. Rev. Lett. \textbf{122} (2019), 172502
doi:10.1103/PhysRevLett.122.172502

\bibitem{Cai:2025txx}
B.~J.~Cai, B.~A.~Li and Y.~G.~Ma,
Eur. Phys. J. Spec. Top. \textbf{2026}
doi:10.1140/epjs/s11734-026-02227-9

\bibitem{Cai:2025mrv}
B.~J.~Cai, B.~A.~Li and Y.~G.~Ma,
Mod. Phys. Lett. A \textbf{0} (2026), 2630005
doi:10.1142/S0217732326300053

\bibitem{Fomin:2026swt}
N.~Fomin, O.~Hen, J.~Kahlbow, \textit{et al.}
[arXiv:2601.09568 [nucl-ex]].

\bibitem{Ye:2024mls}
Z.~Ye, H.~Zhang, Y.~Zhang and H.~Zhao,
Eur. Phys. J. A \textbf{60} (2024), 126
doi:10.1140/epja/s10050-024-01343-1

\bibitem{Subedi:2008zz}
R.~Subedi, R.~Shneor, P.~Monaghan, \textit{et al.}
Science \textbf{320} (2008), 1476-1478
doi:10.1126/science.1156675

\bibitem{Yong:2015gma}
G.~C.~Yong,
Phys. Lett. B \textbf{765} (2017), 104-108
doi:10.1016/j.physletb.2016.12.013

\bibitem{Zhang:2016vcc}
F.~Zhang and G.~C.~Yong,
Eur. Phys. J. A \textbf{52} (2016), 350
doi:10.1140/epja/i2016-16350-4

\bibitem{Reichert:2025egt}
T.~Reichert and J.~Aichelin,
Phys. Lett. B \textbf{880} (2026), 140836
doi:10.1016/j.physletb.2026.140836

\bibitem{Guo:2025xie}
W.~M.~Guo and C.~H.~Chen,
Phys. Rev. C \textbf{111} (2025), 024612
doi:10.1103/PhysRevC.111.024612

\bibitem{Nan:2025xvi}
M.~Z. Nan, P.~C. Li, G.~J. Wei, X.~L. Xiang, W.~Zuo and Q.~F. Li,
[arXiv:2510.09337 [nucl-th]].

\bibitem{Bass:1998ca}
S.~A.~Bass, M.~Belkacem, M.~Bleicher, \textit{et al.}
Prog. Part. Nucl. Phys. \textbf{41} (1998), 255-369
doi:10.1016/S0146-6410(98)00058-1

\bibitem{Bleicher:1999xi}
M.~Bleicher, E.~Zabrodin, C.~Spieles, \textit{et al.}
J. Phys. G \textbf{25} (1999), 1859-1896
doi:10.1088/0954-3899/25/9/308

\bibitem{Li:2011zzp}
Q.~F. Li, C.~W. Shen, C.~C. Guo, Y.~J. Wang, Z.~X. Li, J.~Lukasik and W.~Trautmann,
Phys. Rev. C \textbf{83} (2011), 044617
doi:10.1103/PhysRevC.83.044617

\bibitem{Wang:2020dru}
Y.~J. Wang, Q.~F. Li, Y.~Leifels and A.~Le F{\`e}vre,
Phys. Lett. B \textbf{802} (2020), 135249
doi:10.1016/j.physletb.2020.135249

\bibitem{Li:2022wvu}
P.~C. Li, Y.~J. Wang, Q.~F. Li and H.~F. Zhang,
Phys. Lett. B \textbf{828} (2022), 137019
doi:10.1016/j.physletb.2022.137019

\bibitem{Liu:2020jbg}
Y.~Y. Liu, Y.~J. Wang, Y.~Cui, C.~J.~Xia, Z.~X. Li, Y.~Chen, Q.~F. Li and Y.~X. Zhang,
Phys. Rev. C \textbf{103} (2021), 014616
doi:10.1103/PhysRevC.103.014616

\bibitem{Li:2021thg}
B.~A.~Li, B.~J.~Cai, W.~J.~Xie and N.~B.~Zhang,
Universe \textbf{7} (2021), 182
doi:10.3390/universe7060182

\bibitem{Zhang:2013wna}
Z.~Zhang and L.~W.~Chen,
Phys. Lett. B \textbf{726} (2013), 234-238
doi:10.1016/j.physletb.2013.08.002

\bibitem{Cozma:2021tfu}
M.~D.~Cozma and M.~B.~Tsang,
Eur. Phys. J. A \textbf{57} (2021), 309
doi:10.1140/epja/s10050-021-00616-3

\bibitem{Wang:2018hsw}
Y.~J. Wang, C.~C. Guo, Q.~F. Li, A.~Le F{\`e}vre, Y.~Leifels and W.~Trautmann,
Phys. Lett. B \textbf{778} (2018), 207-212
doi:10.1016/j.physletb.2018.01.035

\bibitem{Russotto:2016ucm}
P.~Russotto, S.~Gannon, S.~Kupny, \textit{et al.}
Phys. Rev. C \textbf{94} (2016), 034608
doi:10.1103/PhysRevC.94.034608

\bibitem{Tong:2020dku}
L.~Y. Tong, P.~C. Li, F.~P. Li, Y.~J. Wang, Q.~F. Li and F.~X. Liu,
Chin. Phys. C \textbf{44} (2020), 074101
doi:10.1088/1674-1137/44/7/074103

\bibitem{Wang:2002ywa}
N.~Wang, Z.~X.~Li and X.~Z.~Wu,
Phys. Rev. C \textbf{65} (2002), 064608
doi:10.1103/PhysRevC.65.064608

\bibitem{Xiang:2025dxt}
X.~L. Xiang, M.~Nan, P.~C. Li, Y.~J. Wang, L.~Liu and Q.~F. Li,
Phys. Rev. C \textbf{113} (2026), 064606
doi:10.1103/ymmt-f8h7

\bibitem{Zhang:2022tsw}
F.~Zhang and G.~C.~Yong,
Phys. Rev. C \textbf{106} (2022), 054603
doi:10.1103/PhysRevC.106.054603

\bibitem{Yang:2018xtl}
Z.~X.~Yang, X.~H.~Fan, G.~C.~Yong and W.~Zuo,
Phys. Rev. C \textbf{98} (2018), 014623
doi:10.1103/PhysRevC.98.014623

\bibitem{Guo:2021zcs}
W.~M.~Guo, B.~A.~Li and G.~C.~Yong,
Phys. Rev. C \textbf{104} (2021), 034603
doi:10.1103/PhysRevC.104.034603

\bibitem{Li:2000sha}
Q.~F.~Li, Z.~X.~Li and G.~J.~Mao,
Phys. Rev. C \textbf{62} (2000), 014606
doi:10.1103/PhysRevC.62.014606

\bibitem{Li:2003vd}
Q.~F.~Li, Z.~X.~Li and E.~G.~Zhao,
Phys. Rev. C \textbf{69} (2004), 017601
doi:10.1103/PhysRevC.69.017601

\bibitem{Li:2017pis}
Q.~F. Li and Z.~X. Li,
Sci. China Phys. Mech. Astron. \textbf{62} (2019), 972011
doi:10.1007/s11433-018-9336-y

\bibitem{Nan:2023gwp}
M.~Z. Nan, P.~C. Li, Y.~J. Wang, Q.~F. Li and W.~Zuo,
Eur. Phys. J. A \textbf{60} (2024), 131
doi:10.1140/epja/s10050-024-01349-9

\bibitem{Nan:2024ogc}
M.~Z. Nan, P.~C. Li, W.~Zuo and Q.~F. Li,
Chin. Phys. C \textbf{49} (2025), 094112
doi:10.1088/1674-1137/add8fd

\bibitem{Li:2006ez}
Q.~F.~Li, Z.~X.~Li, S.~Soff, M.~Bleicher and H.~Stoecker,
J. Phys. G \textbf{32} (2006), 407-416
doi:10.1088/0954-3899/32/4/001

\bibitem{Russotto:2011hq}
P.~Russotto, P.~Z.~Wu, M.~Zoric, \textit{et al.}
Phys. Lett. B \textbf{697} (2011), 471-476
doi:10.1016/j.physletb.2011.02.033

\bibitem{Guo:2024zij}
D.~Guo, X.~H. He, P.~C. Li, \textit{et al.}
Eur. Phys. J. A \textbf{60} (2024), 36
doi:10.1140/epja/s10050-024-01245-2

\bibitem{Miyatsu:2022wuy}
T.~Miyatsu, M.~K.~Cheoun and K.~Saito,
Astrophys. J. \textbf{929} (2022), 82


\bibitem{Sun:2022yor}
X.~Sun, Z.~Miao, B.~Sun and A.~Li,
Astrophys. J. \textbf{942} (2023), 55

\bibitem{Guo:2015tra}
W.~M.~Guo, G.~C.~Yong and W.~Zuo,
Phys. Rev. C \textbf{92} (2015), 054619
doi:10.1103/PhysRevC.92.054619

\bibitem{Serot:1997xg}
B.~D.~Serot and J.~D.~Walecka,
Int. J. Mod. Phys. E \textbf{6} (1997), 515-631
doi:10.1142/S0218301397000299

\bibitem{Furnstahl:1999ff}
R.~J.~Furnstahl and B.~D.~Serot,
Nucl. Phys. A \textbf{673} (2000), 298-310
doi:10.1016/S0375-9474(00)00146-9

\bibitem{Plohl:2006hy}
O.~Plohl and C.~Fuchs,
Phys. Rev. C \textbf{74} (2006), 034325
doi:10.1103/PhysRevC.74.034325


\bibitem{lpcHBT}
P.~C. Li, Y. J. Wang, and Q. F. Li,
[arXiv:2609.23422 [nucl-th]].

\bibitem{hades}
M.~Lorenz, [for the HADES Collaboration]
https://indico.gsi.de/event/22170/contributions/94212/attachments/53458

\bibitem{Arnaud}
A.~Le F{\`e}vre and M.~Lorenz, private communication.

\bibitem{Yong:2015wha}
G.~C.~Yong,
Phys. Rev. C \textbf{93} (2016), 044610
doi:10.1103/PhysRevC.93.044610

\bibitem{HADES:2022osk}
J.~Adamczewski-Musch \textit{et al.} [HADES],
Eur. Phys. J. A \textbf{59} (2023), 80
doi:10.1140/epja/s10050-023-00936-6

\bibitem{Wang:2024ktk}
Y.~J. Wang, B.~Gao, G.~J. Wei, P.~C. Li and Q.~F. Li,
Phys. Rev. C \textbf{110} (2024), 044606
doi:10.1103/PhysRevC.110.044606

\bibitem{Reichert:2024ayg}
T.~Reichert and J.~Aichelin,
Phys. Rev. C \textbf{111} (2025), 054916
doi:10.1103/PhysRevC.111.054916

\end{thebibliography}
\end{document}